\documentclass{ws-ijmpa}
\usepackage[super,compress]{cite}
\usepackage{graphicx}
\usepackage{xspace}
\usepackage{comment}
\usepackage{url}
\usepackage{hyperref}
\hypersetup{colorlinks,citecolor=black,filecolor=black,linkcolor=black,urlcolor=blue}

\newcommand{\dd}{\ensuremath{\mathrm d}\xspace}
\newcommand{\ee}{\ensuremath{\mathrm e}\xspace}

\usepackage{aas_macros}

\usepackage{soul}
\usepackage{multirow}

\usepackage{subfigure}

\usepackage{orcidlink}
\newcommand{\orcidA}{\orcidlink{0000-0003-0328-1226}} 
\newcommand{\orcidB}{\orcidlink{0000-0002-5735-6273}} 
\newcommand{\orcidC}{\orcidlink{0000-0001-9223-6480}} 

\begin{document}
\markboth{A. Horváth, E. Forgács-Dajka, G. G. Barnaföldi}{Compact star structure in multidimensional Kaluza\,--\,Klein space-time}

%
\catchline{}{}{}{}{}
%

\title{The effect of multiple extra dimensions on the maximal mass of compact stars in Kaluza\,--\,Klein space-time}

\author{Anna Horv\'ath\orcidA$^{\ast,\dagger}$}
\address{horvath.anna@wigner.hun-ren.hu}

\author{Emese Forgács-Dajka\orcidB$^{\ast,\ddagger}$}
\address{e.forgacs-dajka@astro.elte.hu}

\author{Gergely Gábor Barnaföldi\orcidC$^{\ast}$}
\address{barnafoldi.gergely@wigner.hun-ren.hu}

\address{$^{\ast}$HUN-REN Wigner Research Centre for Physics, \\ 29--33 Konkoly--Thege Mikl\'os \'ut, H-1121 Budapest, Hungary \\ 
$^{\dagger}$E\"otv\"os Lor\'and University, Institute of Physics and Astronomy, Department of Astronomy, \\  P\'azm\'any P\'eter s\'et\'any 1/A, H-1117 Budapest, Hungary \\
$^{\ddagger}$HUN-REN-SZTE Stellar Astrophysics Research Group, \\  Szegedi út, Kt. 766, H-6500 Baja, Hungary\\}

\maketitle

\begin{abstract}
Compact stars in the Kaluza\,--\,Klein space-time are investigated, with multiple additional compactified spatial dimensions ($d$). Within the extended phenomenological model, a static, spherically symmetric solution is considered, with the equation of state provided by a zero temperature, interacting multi-dimensional Fermi gas. The maximal masses of compact stars are calculated for different model parameters. We investigated the effect of the existence of multiple extra compactified dimensions within the Kaluza\,--\,Klein compact star structure. We found that the number of extra dimensions plays a similar role, and to a similar order, as the excitation number: increasing their number, $d$, reduces the maximal mass by a few percent.

\end{abstract}


\section{Introduction} \label{sec:intro}

The study of neutron stars and other compact astrophysical objects offers a unique opportunity to understand particle physics and the extreme conditions in the Universe. The high densities of these objects and the extreme conditions of their inner matter make them ideal laboratories for testing hypotheses beyond the Standard Model of particle physics, including exotic particles and extra dimensions proposed by e.g. the Kaluza\,--\,Klein theory.

Neutron stars are formed as the remnants of gravitational collapse supernovae, occurring at the end of the life cycle of stars whose initial mass exceeds 5 solar masses~\cite{2012PTEP.2012aA309J, 2021Natur.589...29B}. During various stages of stellar evolution, the star loses significant mass, partly due to stellar winds and partly as a result of the supernova explosion. The collapsing stellar core, having lost equilibrium, reaches nuclear matter density, and thus the nuclear forces become repulsive. Initially, the proto-neutron star is lepton-rich, 
with high opacity of the core. However, after a few seconds, approximately 99\% of the neutron star's binding energy is radiated away as neutrinos~\cite{1986ApJ...307..178B}. The core quickly cools, and within minutes, thermal processes no longer  affect significantly its structure and composition, except for the emergence of superfluidity. 

Five main regions can be identified in a neutron star: the atmosphere, the envelope, the crust, the outer core, and the inner core~\cite{2010NewAR..54..101L}. The neutron star's atmosphere is a very thin layer, only a few centimeters thick, which determines the shape of the star's observable X-ray spectrum. The envelope is about 10 to 100 meters thick.
In the neutron star's crust, as density increases, the nuclei become increasingly neutron-rich, and in the densest regions, they might deform, 
creatig so-called "pasta" phases~\cite{2024PhRvD.109f3022M, 2021A&A...654A.114D}. Finally, at the boundary of the outer core, a phase transition into homogeneous nuclear matter may occur. It is not clear if a distinct boundary exists between the outer and inner core. Latter's exotic composition is subject to various predictions~\cite{2006ARNPS..56..327P}: hyperons, condensed kaons or pions, or even deconfined quark matter, either in a pure phase, or mixed with hadrons. Decofinement means that quarks and gluons are no longer bound within protons, neutrons and other hadrons, but exist freely in what is known as quark-gluon plasma~\cite{2010PhRvD..81j5021K}.

The constraints of causality, the observed maximum mass of neutron stars, and recent observations suggesting a neutron star radius smaller than 13.5~km, place strong limits on the parameter space for models predicting exotic particle production. The high density of the material surrounding the core also suppresses observable effects of deconfinement, but there may be events or dynamic phenomena in the life of the neutron star that could allow for their detection. Such events include neutron star mergers, where tidal resonances or distortions observed in various oscillation modes may carry detectable gravitational wave signals.

In summary, neutron stars and related objects (such as magnetars, quark stars etc.) are not only of astrophysical significance, but also serve as laboratories for probing the shortcomings of the Standard Model and addressing deeper cosmological questions, or even to test the existence of extra dimensions and exotic particles~\cite{Lukacs:2003fh,Barnafoldi:2007,Barnafoldi:2015wca,Barnafoldi:2020phq}. This paper follows the idea proposed by Horváth {\it et al.}~\cite{Horvath_2024} for one extra compactifed dimension. It aims to explore how the existence of multiple extra dimensions modifies the equations governing matter and how these modifications at the microscopic, or nuclear scale affect the macroscopic properties of compact objects. 


\section{Compact stars in Kaluza~--~Klein space-time}
 \label{sec:KK}

To derive observable quantities on an astronomical scale, it is necessary to link the microscopic, nuclear physics properties of matter with the structure of compact objects. The Kaluza\,--\,Klein-like multidimensional model, which we are investigating here, influences physics at the subnuclear level, leading to modifications in the equations that describe matter, and through statistical physics, it manifests in the equation of state. Compact astrophysical objects within the general relativistic framework are described by the Tolman\,--\,Oppenheimer\,--\,Volkoff (TOV) equation, which models stars in hydrostatic equilibrium with relativistic corrections in a static, spherically symmetric spacetime. The TOV equation provides the pressure change with respect to the star's radius at each point, using the local energy density, thereby incorporating thermodynamics (EoS) into the star's structure. In this section, we outline the fundamental components of this process, then integrate them to demonstrate how the macroscopic properties of compact objects are computed.

\subsection{The Kaluza~--~Klein space-time and geometry with multiple extra dimensions}
\label{sec:kkspacetime}

Theodor Kaluza introduced a geometric interpretation for electromagnetism, integrating it with gravity within a five-dimensional metric tensor as detailed in Ref.~\citen{Kaluza:1921tu}. A pivotal aspect of this framework is the implementation of the "cylinder condition", which stipulates that the metric components are independent of the fifth coordinate\footnote{Traditionally the 4\textsuperscript{th} spatial coordinate is denoted by index "5" for being of the fifth compactified dimension.}. Subsequently, Oskar Klein proposed that the fifth dimension should be compactified and microscopic, as discussed in his famous work~\cite{Klein:1926tv}. This compactification provides a quantum-mechanical motivation for the extra dimension's negligible impact on macroscopic scales. According to this interpretation, spacetime comprises the conventional one temporal and three infinite, large-scale spatial dimensions, augmented by an additional compactified microscopic spatial dimension, which is curled into a circle at every point in the usual space.
\begin{figure}[!ht]
     \centering
     \includegraphics[scale=0.35]{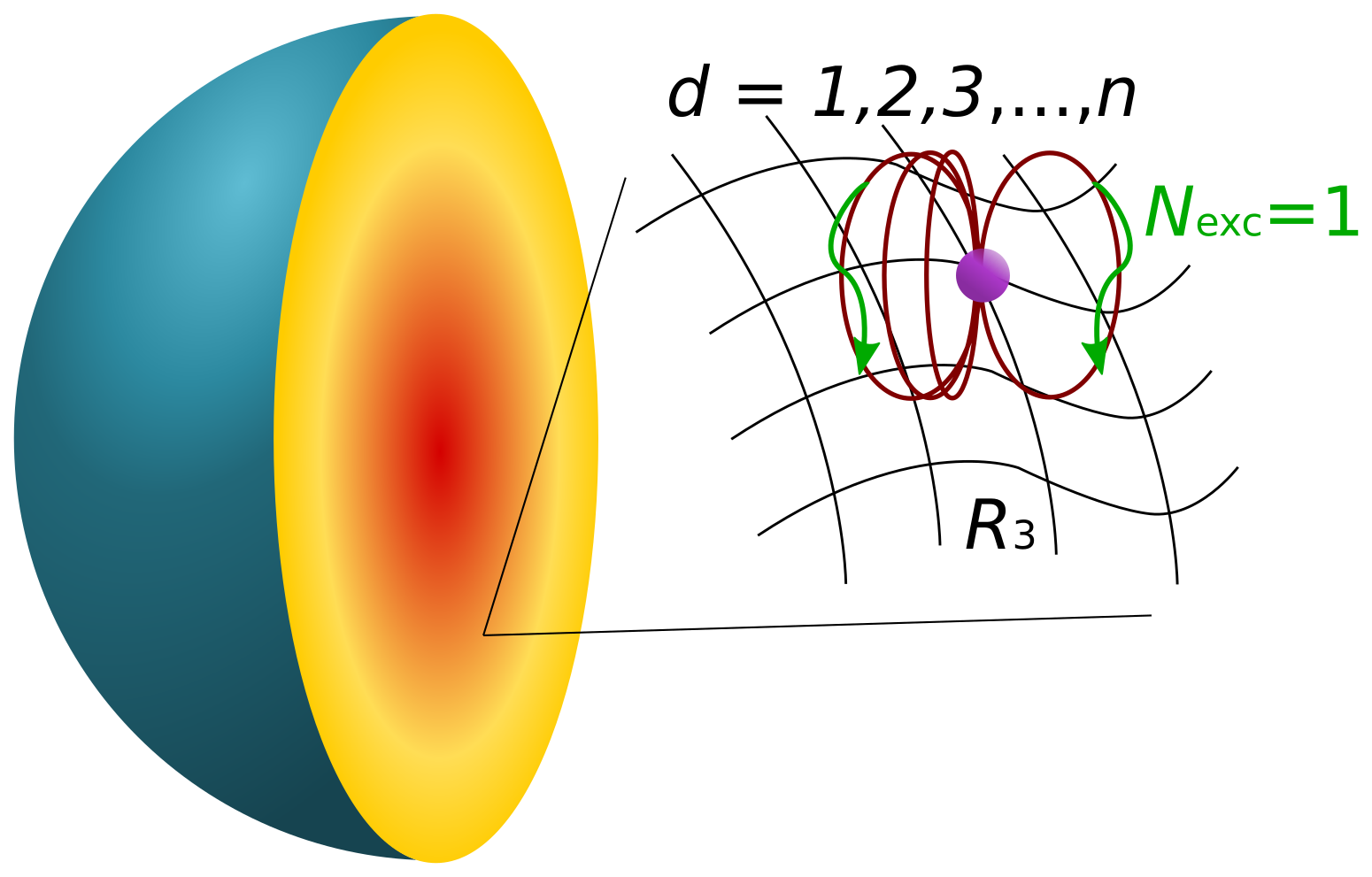}
     \caption{Schematic view of the 3 infinite spacial dimensions ($R^3$) and the $d$ compactified extra spatial ones influencing physics in the extreme environment of a compact astrophysical object's interior. Having high-enough energy, a particle (purple ball) can acquire excitations (green wavy arrows) in the directions of some, or all of the extra dimensions.}
     \label{fig:scheme}
\end{figure}

As an extension of the original idea of Kaluza and Klein, in this study we investigate a model with not only one, but multiple-added extra compactified spatial dimensions, existing at each point in usual 1+3 dimensional spacetime (see Fig.~\ref{fig:scheme}). This type of topology is common among beyond standard model theories, such as brane- and string theory. According to our assumption, standard model particles can propagate in these extra dimensions, and, due to their finite size (compactification radii), they acquire quantized momenta in them. In the current work, we neglect the effects of electromagnetism, which traditionally appears in the Kaluza\,--\,Klein theory, and consider a static, spherically symmetric spacetime, thus we can set the off-diagonal elements of the metric tensor to zero: 
\begin{equation}
    g_{AB}=\mathrm{diag} \left[\begin{matrix}g_{00}, g_{11}, g_{22}, g_{22} \sin^{2}\theta,  g_{55},\dots ,g_{dd}\end{matrix}\right]\ , 
    \label{eq:diagmetric}
\end{equation}
where $A$ and $B$ are indices running from 0 to $3+d$, $d$ is the number of extra compactified spatial dimensions and $\theta$ is the polar angle. The extra-dimensional components of the metric can be interpreted geometrically, but also in a field theoretical way, considering the appearance of scalar fields.

One can express the energy of a particle moving in 1+3+$d$ dimensional spacetime using natural units ($c=1=\hbar$) as
\begin{equation}
    E^2 = \mathbf{k}^2+ \tilde{k}^2_1 + \dots +\tilde{k}^2_d + m^2 = \mathbf{k}^2+\sum\limits_{j=1}^{d} \tilde{k}_j^2+m^2 = \mathbf{k}^2+\sum\limits_{j=1}^{d} \Big(\frac{n_j}{r_{{\mathrm{c}}_j}}\Big)^2+m^2\ ,
    \label{eq:energy}
\end{equation}
where $m$ is the rest mass of the particle in 1+3+$d$ dimensions, $\mathbf{k}$ is the standard, 3-dimensional momentum, $\tilde{k}_j =n_j/r_{\mathrm{c}_j} $ are the momenta in the extra dimensions, $N_\mathrm{exc}$ is the maximal excitation number, $n_j \in \{ 0,...,N_{\mathrm{exc}} \} $ and the $r_{\mathrm{c}_j}$ are the radii of the extra compactified spatial dimensions. The momentum components of Eq.~\eqref{eq:energy} play a similar role as the mass, thus term $ \sum_{j=1}^{d} \tilde{k}_j^2 $ can be interpreted as a contribution to the particle's mass in the usual picture, where spacetime is 1+3-dimensional:
\begin{equation}
    \label{eq:meff}
    \tilde{m}^2 = m^2 + \sum_{j=1}^{d}\tilde{k}_j^2\ ,
\end{equation}
producing a 1+3 dimensional effective mass, $\tilde{m}$.

\subsection{The equation of state of the multidinemsional Fermi gas}
\label{sec:eos}

A Kaluza–Klein-like theory can be linked to the macroscopic observables of a compact object by its impact on thermodynamics. In our approach, we modeled the matter inside a neutron star using the equation of state (EoS) for a Fermi gas. However, when considering the extra-dimensional framework, we must extend the model to incorporate more than three spatial dimensions. The thermodynamic potential in $3$ spatial and additional $d$ compactified extra dimensions can be written as
\begin{eqnarray}
    \label{eq:termpot}
    \Omega =&&- V_{(3+d)}\sum\limits_{i}\underbrace{\sum\limits_{j=0}^{N_\mathrm{exc}} \dots \sum\limits_{l=0}^{N_\mathrm{exc}}}_{d} \frac{g_i}{\beta}\int\limits \frac{\dd^{3} \textbf{k}}{(2\pi)^{3}} \times   \\ 
    && \hspace{3truecm} \times  \left[\ln \left(1+\ee^{-\beta(E_{ij \dots l}-\mu)}\right)+\ln \left(1+\ee^{-\beta(E_{ij \dots l}+\mu)}\right) \right]\ , \nonumber
\end{eqnarray}
with a summation over different fermions, $i \in \{n^0,p^+,\dots \}$, with one half-spin of multiplicity $g_i =2$. The $3+d$-dimensional volume is denoted by $V_{(3+d)}$, $\beta=1/k_BT$ is the inverse temperature, and the 3-momentum in the usual three infinite dimensions is $\mathbf{k}$. The energy of particle $i$ with excitations $j \dots l$ in the extra compactified dimensions is $E_{ij \dots l}$ and $\mu_i$ is the chemical potential.

Summation over the $j \dots l$ indices corresponds to the excitations in the extra compactified dimensions. This produces $(N_{\mathrm{exc}}+1)^d$ different states, since we consider the extra compactified dimensions to be distinguishable. However, we also set the size of each extra dimension to be the same ($r_{\mathrm{c}_j}=r_{\mathrm{c}}$ for all $j$), and assume that all the maximal excitation numbers, $N_\mathrm{exc}$ are equal. Thus the energies of some of these states will be the same. Reducing the theory to the usual 1+3 dimensional case that we can observe, these energy states can be interpreted as particles having effective mass, $\bar{m}$, as specified by Eq.~\eqref{eq:meff}, indeed all the multiplicities can be inserted to a factor, $\tilde{g}_{ij}$. Thus one can simply consider a thermodynamic potential with a summation over these effective mass states with different multiplicities depending on the number of extra compactified dimension and the given state:  
\begin{equation}
    \label{eq:termpot2}
    \Omega=-V_{(3+d)}\sum\limits_{i}\sum\limits_{j=0}^{N_\mathrm{exc}\cdot d} \frac{\Tilde{g}_{ij}}{\beta}\int\limits \frac{\dd^{3} \textbf{k}}{(2\pi)^{3}} \times  \left[\ln \left(1+\ee^{-\beta(E_{ij}-\mu)}\right)+\ln \left(1+\ee^{-\beta(E_{ij}+\mu)}\right) \right]\ .
\end{equation}

In the current work, we take the zero temperature limit of the equation of state, since in neutron stars the typical temperature values are negligible in comparison to the nuclear bounding energy. 
The thermodynamic variables, such as pressure, $p$, energy density, $\varepsilon$, chemical potential, $\mu$ and baryon number density, $n$ can be derived as discussed in our previous works~\cite{Barnafoldi:2007,Horvath_2024}. An important aspect of the equation of state, however, is that we model the short-range (nuclear) force acting between particles using a simple potential linear in the number density, 
\begin{equation}
    U(n) = \xi n \ ,
    \label{eq:poti}
\end{equation}
where $\xi$  is a constant parameter corresponding to the strength of the interaction with the typical value of $\sim {\cal O} (1)$~GeV$\cdot$fm\textsuperscript{3}, which can be associated with $ g^2_{\omega}/m^2_{\omega}$ following from nuclear  theory~\cite{Zimanyi:1987bt}.  This potential is repulsive, which is a good model at the high density regime. Thermodynamical variables will be modified due to this interaction through the chemical potential and an interaction term as
\begin{equation}
    \begin{aligned}
        p(\mu)&=p_0(\bar\mu)+p_{\mathrm{int}} \\
        \varepsilon(\mu)&=\varepsilon_0(\bar\mu)+\varepsilon_{\mathrm{int}}\ ,
        \label{eq:ep}
    \end{aligned}
\end{equation}
where $p_0$ and $\varepsilon_0$ correspond to the pressure and energy density of the free theory, $\bar{\mu}=\mu-U(n)$, while the interaction terms are simply the potential integrated with respect to the number density:
\begin{equation}
p_{int}=\varepsilon_{int}=\int U(n)dn = \frac{1}{2}\xi n^2 \ ,
\end{equation}
which can be connected directly to the multi-dimensional Fermi gas model, since the calculation of the $n$ is straightforward.

\subsection{Calculating the equation of state of a Kaluza~--~Klein space-time with multiple compactified dimensions}

We examine the properties of the derived equation of state (EoS) in comparison with various selected models of the compact star interior, such as those from Ref.~\citen{compose2}. 
\begin{figure}[!ht]
      \centering
      \hfill
      \begin{subfigure}
         \centering
          \includegraphics[scale=0.38]{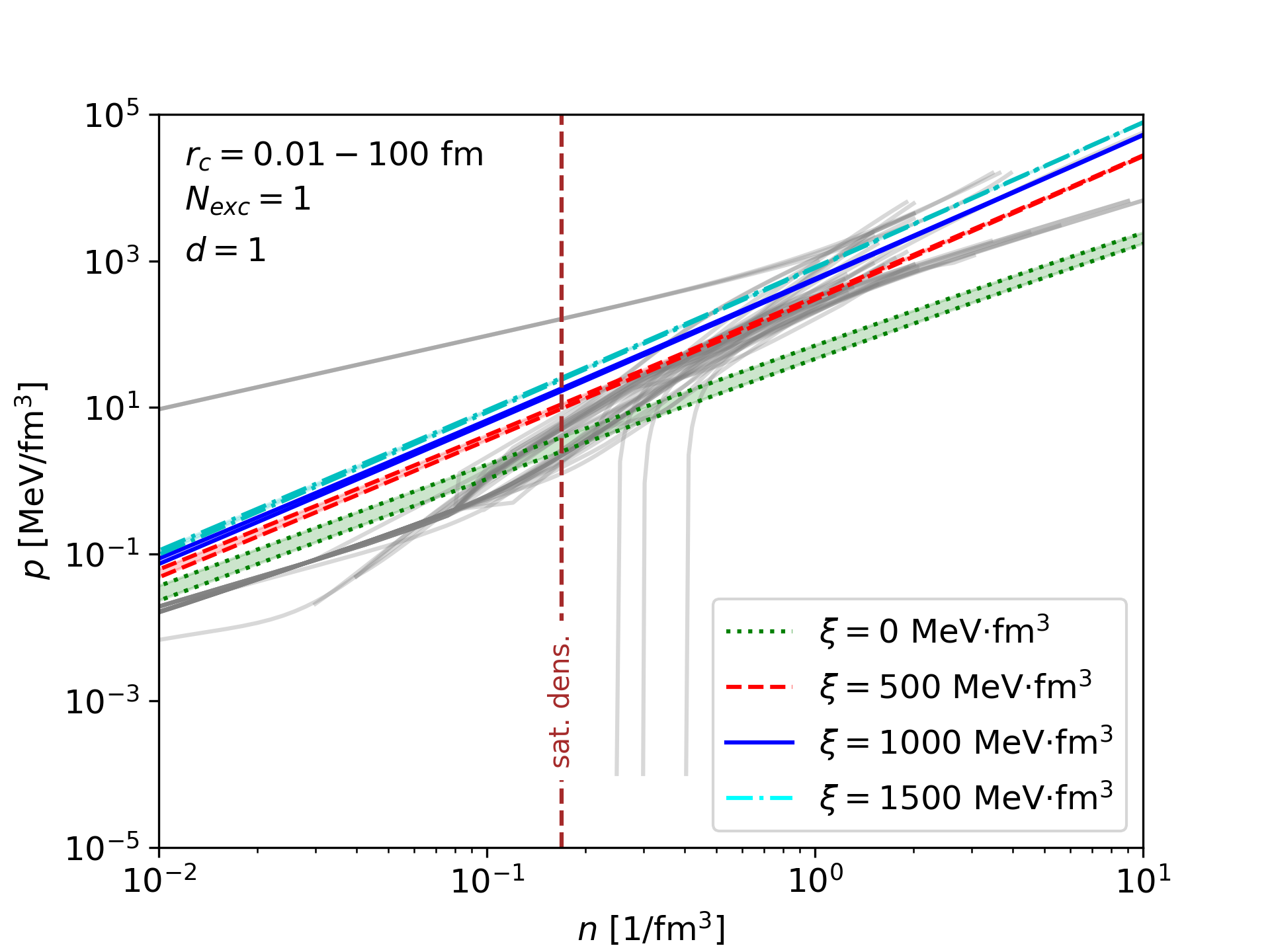}
      \end{subfigure}
      \hfill
      \begin{subfigure}
          \centering
          \includegraphics[scale=0.38]{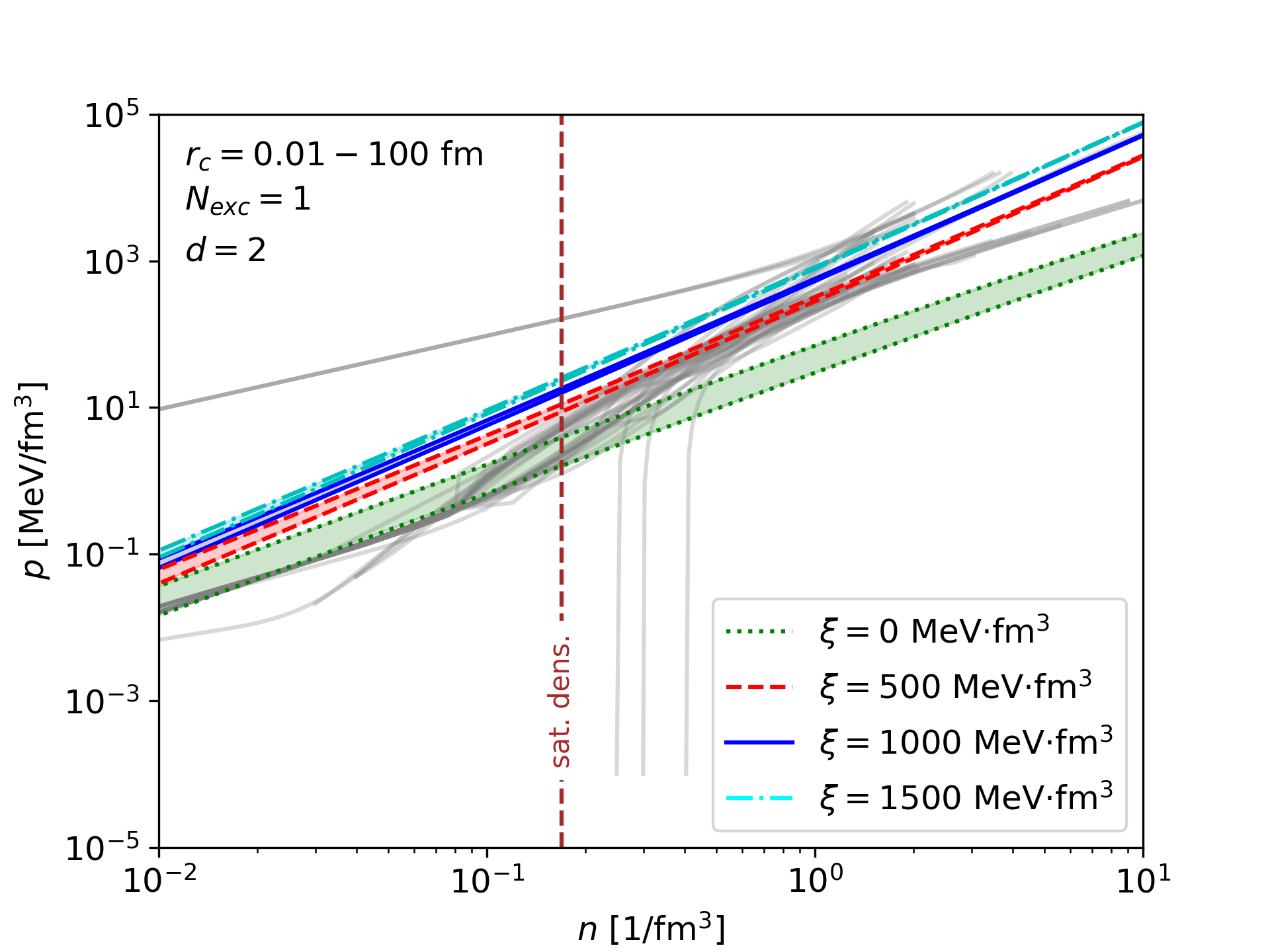}
      \end{subfigure}     
      \hfill

      \hfill
      \begin{subfigure}
          \centering
          \includegraphics[scale=0.38]{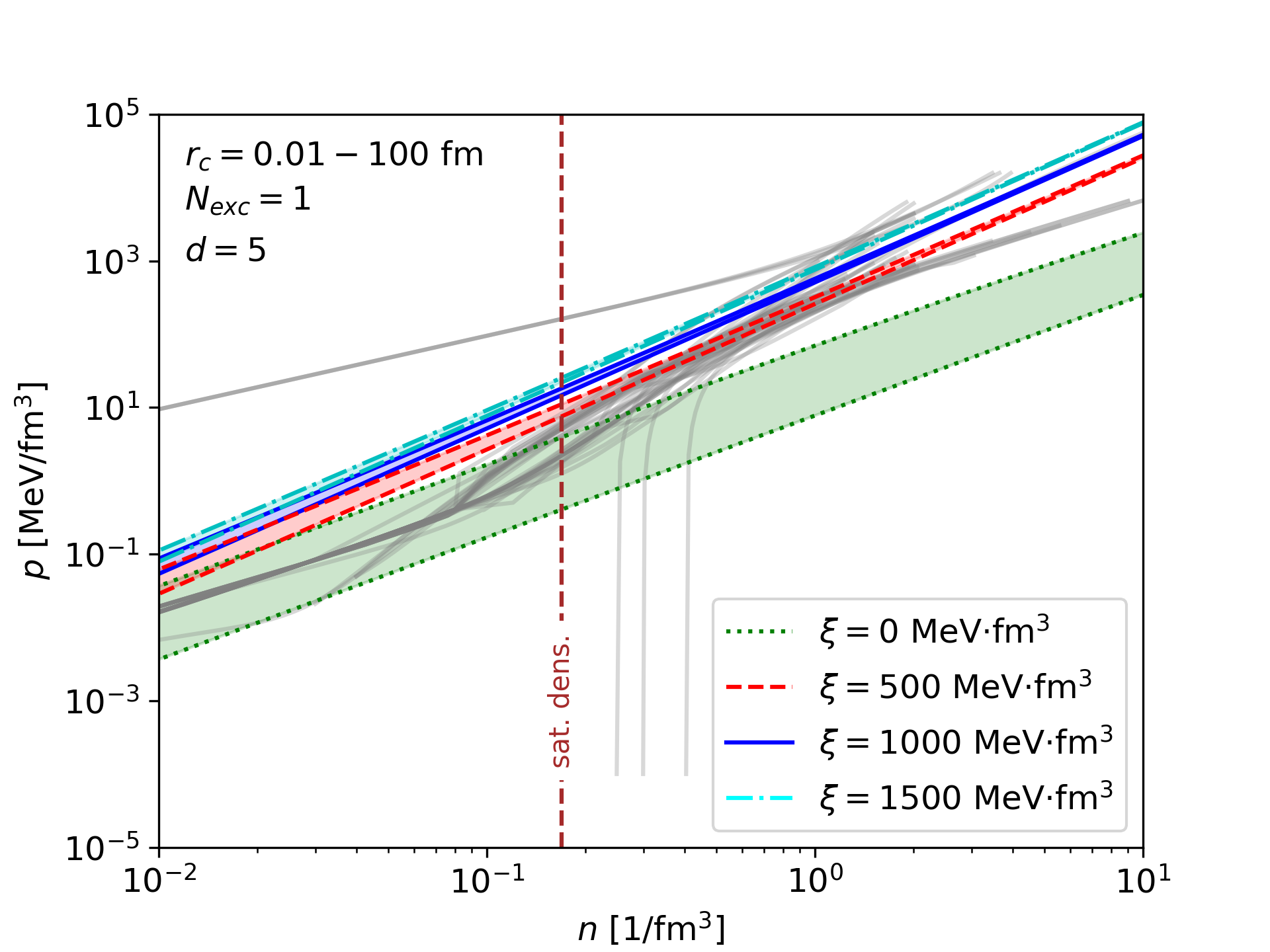}
      \end{subfigure}
      \hfill
      \begin{subfigure}
          \centering
          \includegraphics[scale=0.38]{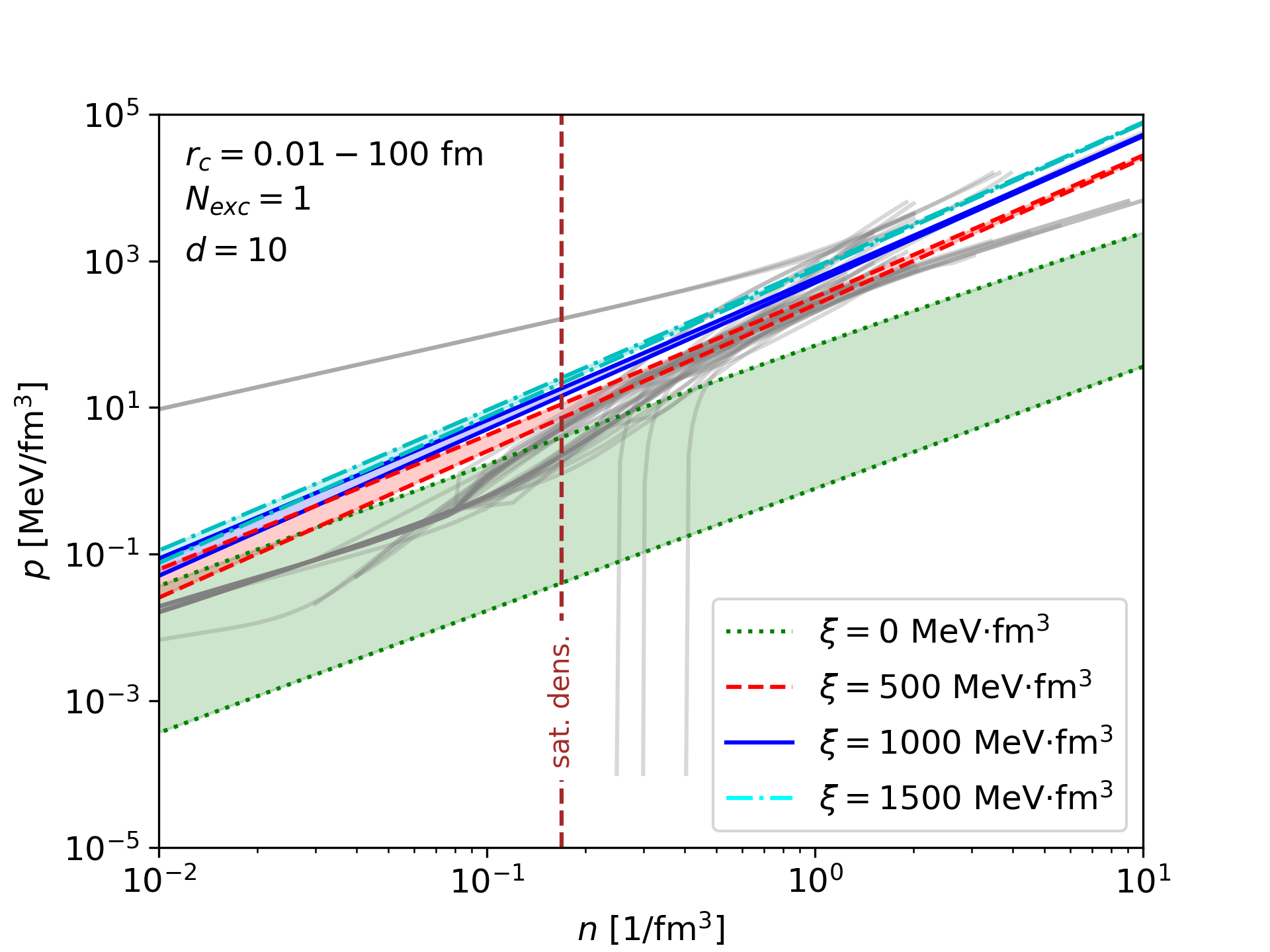}
          
      \end{subfigure}     
      \hfill
    \caption{Equation of state of the multidimensional interacting Kaluza\,--\,Klein gas: pressure, $p$ as a function of number density, $n$, for different numbers of extra compactified spatial dimensions. Different interaction strengths, $\xi$ are indicated by different colors and line styles. Two $r_c$ values, 0.01~fm (upper limit) and 100~fm (lower limit) span bands for each color. We also show the nuclear saturation density with a dashed brown vertical line. For comparison, EoSs of other models are plotted in light gray curves~\cite{compose2,2001ApJ...550..426L, PhysRevD.79.124032, Ozel:2016oaf, PhysRevD.73.024021, Alford_2005, PhysRevC.58.1804, 1997A&A...328..274B, BALBERG1997435, PhysRevC.60.024605, 2013A&A...560A..48P,1996ApJ...469..794E, FRIEDMAN1981502, 1985ApJ...293..470G, PhysRevC.60.025803, MUTHER1987469, MULLER1996508, PhysRevLett.61.2518, PhysRevD.52.661, Douchin:2001sv, PhysRevC.38.1010,RIKOVSKASTONE2007341}.}
      \label{fig:log}
\end{figure}

Pressure, $p$ as a function of number density, $n$ is plotted in Figure~\ref{fig:log}. All "species" of baryons, with different excitation numbers in the different dimensions are summed up to produce $n$. Our calculations span bands on the graph, with upper and lower curves corresponding to extra dimension sizes of $r_c=0.01$~fm and $r_c=100$~fm, respectively. For simplicity, in all four graphs, the maximal possible excitation number was set to $N_\mathrm{exc}=1$. We changed the number of extra spacial dimensions in the subfigures with $d=1$, 2, 5 and 10. The nuclear saturation density is shown in a vertical brown dashed line. The effect of changing $\xi$ is significant on the equation of state, however, extra dimension also play an important role, especially in the absence of interaction. As the number of extra dimensions increases, and new degrees of freedom become available, bands "open up" at the large $r_{\mathrm{c}}$ limit, where a new excitation requires less energy than at small $r_{\mathrm{c}}$ values. We compare our model to other theories, by plotting their equations of state in light gray curves~\cite{2001ApJ...550..426L, PhysRevD.79.124032, Ozel:2016oaf, PhysRevD.73.024021, Alford_2005, PhysRevC.58.1804, 1997A&A...328..274B, BALBERG1997435, PhysRevC.60.024605, 2013A&A...560A..48P,1996ApJ...469..794E, FRIEDMAN1981502, 1985ApJ...293..470G, PhysRevC.60.025803, MUTHER1987469, MULLER1996508, PhysRevLett.61.2518, PhysRevD.52.661, Douchin:2001sv, PhysRevC.38.1010,RIKOVSKASTONE2007341}.

\begin{figure}[!ht]
      \centering
      \hfill
      \begin{subfigure}
         \centering
          \includegraphics[scale=0.38]{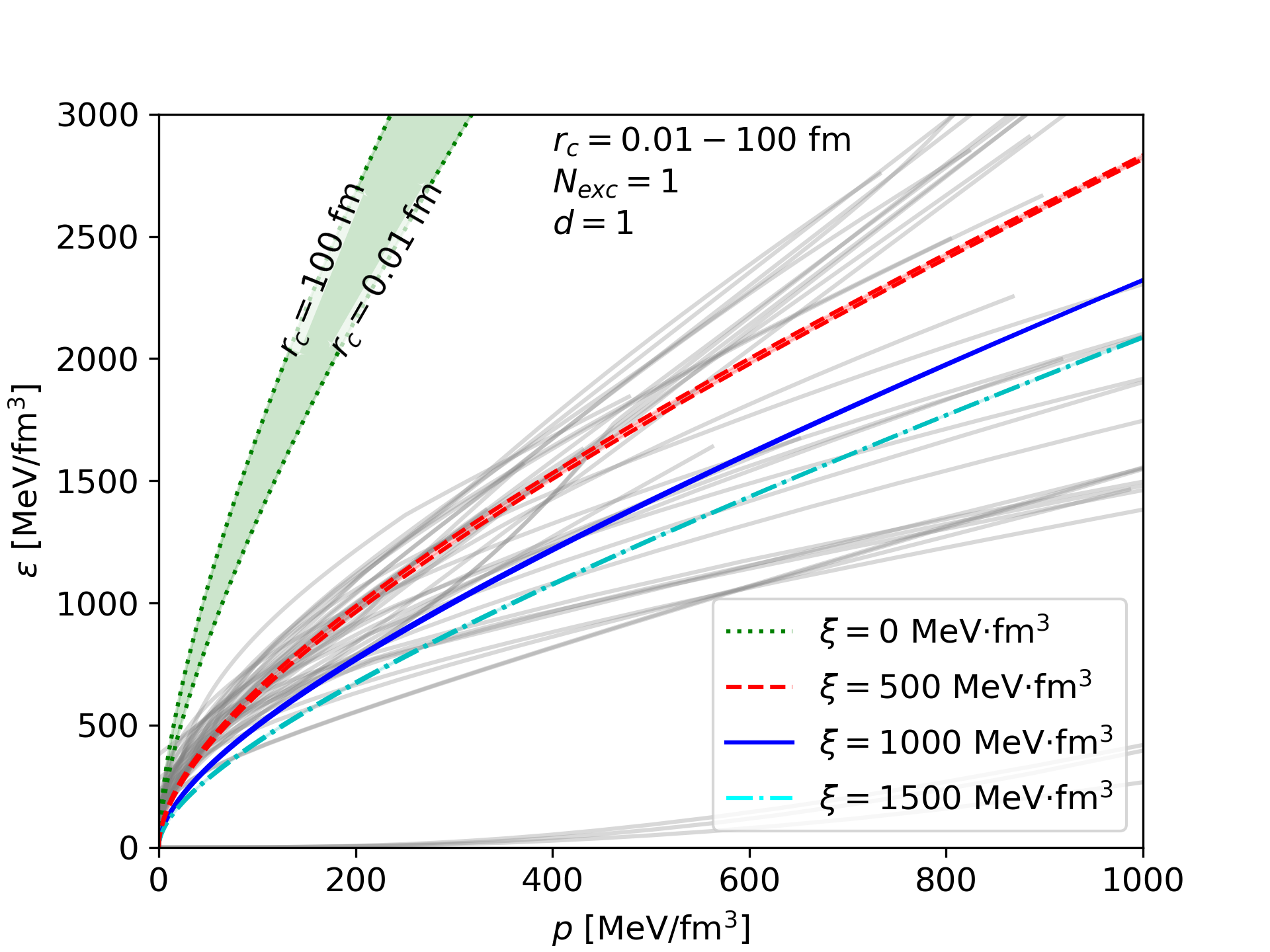}
      \end{subfigure}
      \hfill
      \begin{subfigure}
          \centering
          \includegraphics[scale=0.38]{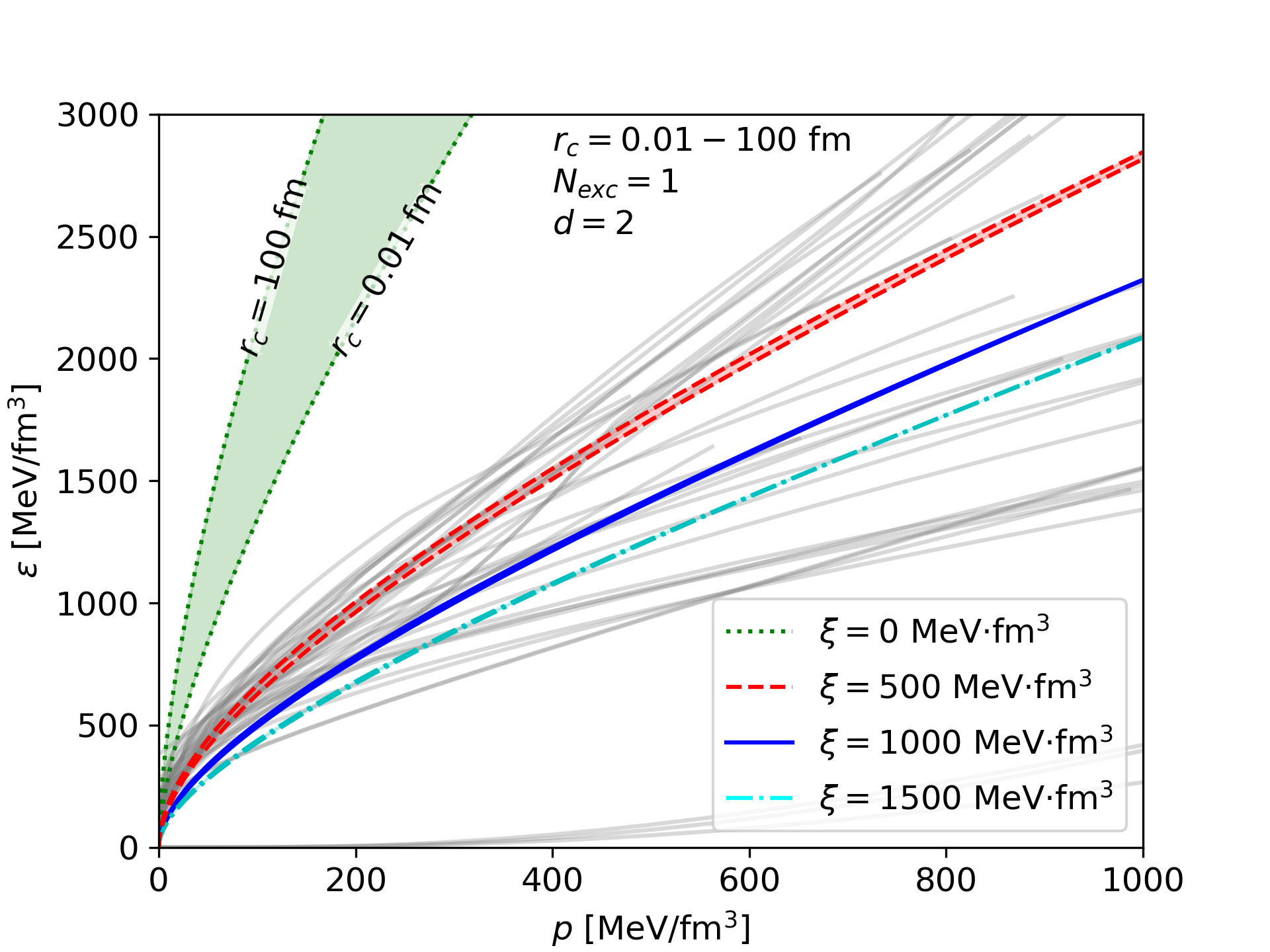}
      \end{subfigure}     
      \hfill

      \hfill
      \begin{subfigure}
          \centering
          \includegraphics[scale=0.38]{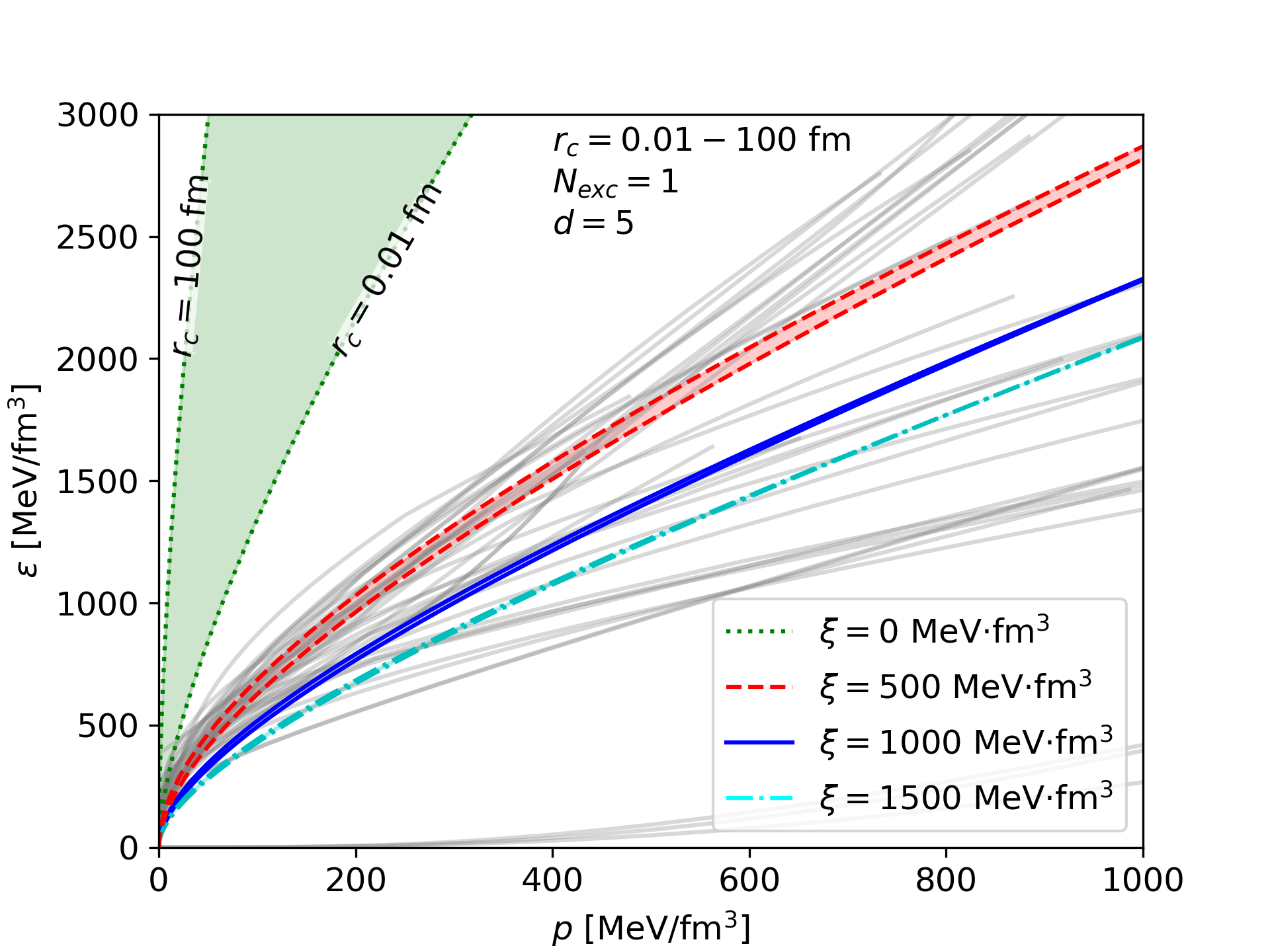}
      \end{subfigure}
      \hfill
      \begin{subfigure}
          \centering
          \includegraphics[scale=0.38]{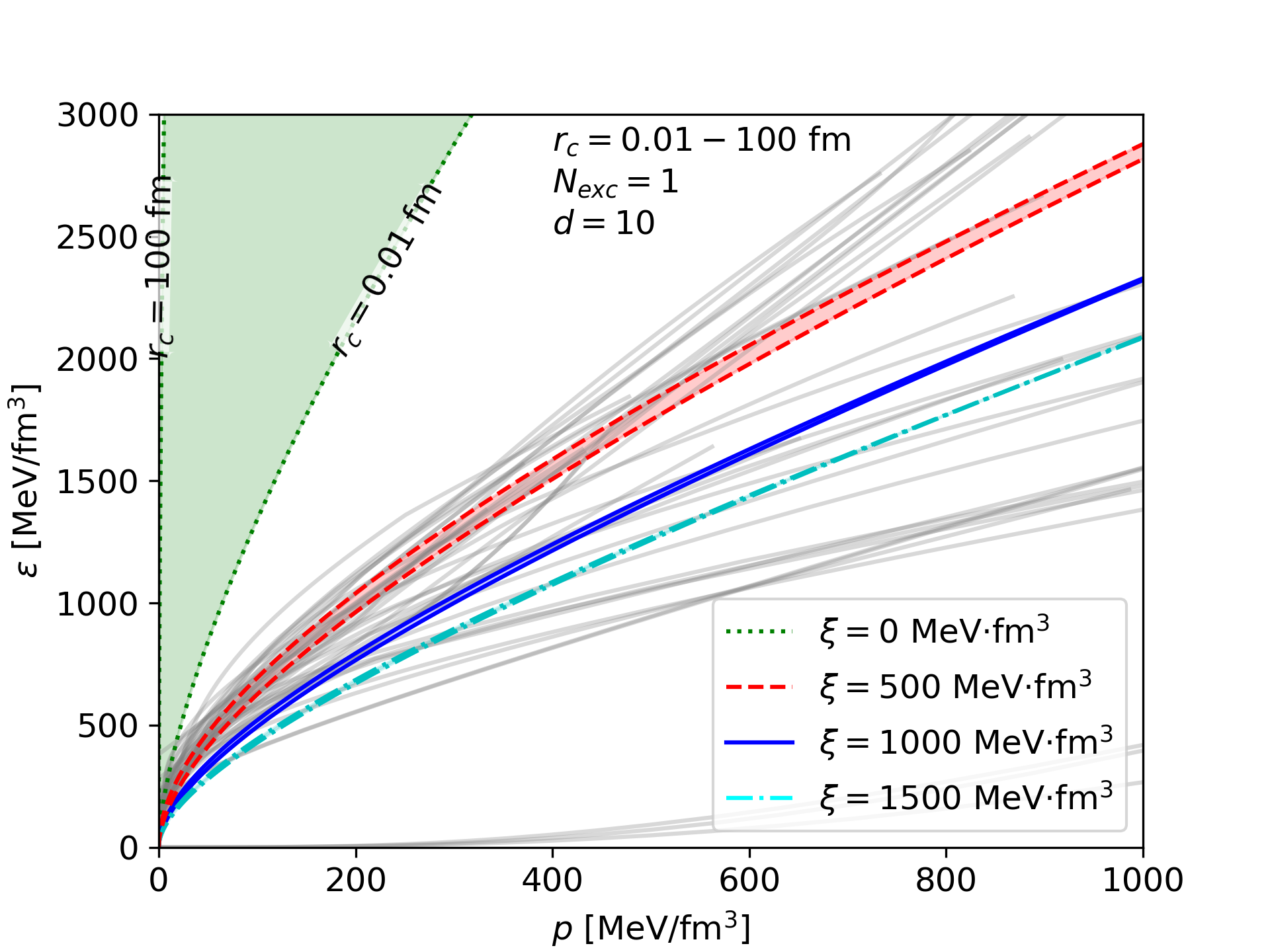}
      \end{subfigure}     
      \hfill

\caption{Equation of state of the multidimensional interacting Kaluza\,--\,Klein gas: energy density, $\varepsilon$ as a function of pressure, $p$. Different interaction strengths, $\xi$ are indicated by different colors and line styles, while $r_{\mathrm{c}}$ values span bands between 0.01 and 100~fm. Panels show EoSs for different extra dimension numbers, $d=1$, 2, 5, and 10. The maximal excitation number, $N_{\mathrm{exc}}=$ 1. We show EoSs taken from~\cite{compose2,2001ApJ...550..426L, PhysRevD.79.124032, Ozel:2016oaf, PhysRevD.73.024021, Alford_2005, PhysRevC.58.1804, 1997A&A...328..274B, BALBERG1997435, PhysRevC.60.024605, 2013A&A...560A..48P,1996ApJ...469..794E, FRIEDMAN1981502, 1985ApJ...293..470G, PhysRevC.60.025803, MUTHER1987469, MULLER1996508, PhysRevLett.61.2518, PhysRevD.52.661, Douchin:2001sv, PhysRevC.38.1010,RIKOVSKASTONE2007341} for comparison (light gray curves).}
      \label{fig:lin}
\end{figure}

In Figure~\ref{fig:lin}, the energy density, $\varepsilon$ is shown as a function of pressure, $p$. Different colors and line styles correspond to different $\xi$ values, as before, while the colored bands are spanned by $r_{\mathrm{c}}$ values between 0.01 and 100~fm. Subfigures differ in the number of extra dimensions: $d=1$, 2, 5 and 10. Here, again the maximal excitation number was set to $N_{\mathrm{exc}}=1$, for simplicity. Gray curves are the equations of state of other theories.~\cite{compose2,2001ApJ...550..426L, PhysRevD.79.124032, Ozel:2016oaf, PhysRevD.73.024021, Alford_2005, PhysRevC.58.1804, 1997A&A...328..274B, BALBERG1997435, PhysRevC.60.024605, 2013A&A...560A..48P,1996ApJ...469..794E, FRIEDMAN1981502, 1985ApJ...293..470G, PhysRevC.60.025803, MUTHER1987469, MULLER1996508, PhysRevLett.61.2518, PhysRevD.52.661, Douchin:2001sv, PhysRevC.38.1010,RIKOVSKASTONE2007341} Similarly to Fig.~\ref{fig:log}, the effect of changing the repulsive coupling $\xi$ is the most prominent, and the larger its value, the less important the extra dimension's effect is. Nonetheless, as more and more states become available by adding extra dimensions in the different panels, the bands, especially those corresponding the smaller $\xi$ values, open up significantly. One can also see that increasing the number and the size of the extra dimension has the opposite effect as increasing the value of $\xi$. In other words, a larger $\xi$ makes the equation of state harder, since an increase in energy density corresponds to a more significant increase in pressure. On the contrary, increasing the number and the size of the extra dimensions softens the EoS. This behavior can easily be understood, since $\xi$ in Eq.~\eqref{eq:poti} represents the repulsive interaction between particles, which makes it harder to fit more into the same volume. However, increasing the number of extra dimensions adds more available states, and with a larger $r_{\mathrm{c}}$, these states become more easily available. 

One can see that these simple Kaluza\,--\,Klein equations of state are similar to those predicted by other models (light gray curves). Above the saturation density, in Fig.~\ref{fig:log}, those are a better approximation for which $\xi$ is nonzero, however, at smaller energies, the green band without interaction seems to be closer to other theories. The explanation for this is that our model's interaction approximates the repulsive regime of the nuclear force, which acts at large densities. Thus, for densities below about the saturation density, which are relevant near the surface and the atmosphere of the star, these equations of state are less accurate.

Adding multiple extra compactified spatial dimensions to spacetime has a similar effect as increasing the maximal excitation number, $N_\mathrm{exc}$, investigated in a previous publication~\cite{Horvath_2024}. Just like for a changing $N_\mathrm{exc}$, new degrees of freedom appear, making the EoSs even softer. For the case studied here, where the $r_\mathrm{c}$ size of all the extra dimensions is the same, even the energy levels are equivalent. However, they appear with different multiplicities, as discussed in Section~\ref{sec:eos}. The effect of adding extra dimensions is stronger, than increasing the excitation number, $N_\mathrm{exc}$, since the number of states is exponential in $d$. Increasing the size of the extra dimensions also makes the EoS softer, however, its effect is qualitatively different from that of $d$ and $N_\mathrm{exc}$. It defines a threshold energy of the system, where new states become available, and where any kind of measurable consequences can be expected. After this threshold is reached, a larger $r_\mathrm{c}$ enhances the effects of $d$ and $N_\mathrm{exc}$. The strongest influence belongs to $\xi$ among the four parameters of the theory, since it can be connected to the nuclear force that is measurable in normal conditions.

\section{Results on the multidimensional Kaluza~--~Klein compact star 
\label{sec:result}} \label{sec:tov}

The equation of state defines the properties of matter, while the Tolman\,--\,Oppenheimer\,--\,Volkoff (TOV) equation describes the equilibrium between gravity and pressure within a static, spherically symmetric, and time-independent object. This remains true in Kaluza\,--\,Klein spacetime as well, as presented by Lukács \& Barnaföldi {\it et al.}~\cite{Lukacs:2003fh,Barnafoldi:2007}, where they demonstrated how the solution for the 5-dimensional case can be reduced to the familiar 4-dimensional TOV equation:
\begin{equation}
\label{eq:tov}
\frac{\dd p(r)}{\dd r} = - \frac{GM(r) \varepsilon(r)}{r^2} \times 
\left[ 1+ \frac{p(r)}{\varepsilon(r)} \right]
\left[ 1+ \frac{4 \pi r^3 p(r)}{M(r)} \right]
\left[ 1 - \frac{GM(r)}{r} \right]^{-1} \ ,
\end{equation}
where $r$ is the radial coordinate, $G$ is the gravitational constant, and $M(r)$ is the mass function
\begin{equation}
\label{eq:mass}
M(r)= \int\limits_{0}^{r} \dd r' 4\pi r'^2 \varepsilon(r') \ .
\end{equation}
The first term in Eq.~\eqref{eq:tov} represents the Newtonian description of hydrostatic equilibrium, while the last three factors account for corrections introduced by general relativity. The generalization from $1+3$ to $1+3+d_\mathrm{c}$ dimensions is analogous to how it is done for $1+3+1_\mathrm{c}$. Esentially, we look for a generalized Schwarzschild solution, where we set the extra $g_{dd}$ components of the  metric~\eqref{eq:diagmetric} to a constant. Eventually, we want a reduced, 1+3 dimensional solution, and it turns out that the microscopic spatial dimensions do not have direct macroscopic effect on the geometry, but affect indirectly the properties of the extreme matter in the interior.

Thus, the TOV, an ordinary differential equation is obtained, that can be solved numerically by integrating with respect to the star's radius, $r$. The equation of state (EoS) establishes the relationship between the energy density, $\varepsilon(r)$ and the pressure, $p(r)$ at each step. To begin, initial boundary conditions are required: we specify the central energy density ($\varepsilon(r=0) = \varepsilon_c$) and assume zero pressure at the surface, $p(r=R)=0$. Starting from the center, the star is constructed layer by layer until a realistic minimum pressure is reached as it approaches zero. Varying the initial condition, $\varepsilon_c$, the mass-radius ($M$-$R$) relation can be calculated for specific cases. In this study we focused on the effect of the size ($r_{\mathrm{c}}$) and the number of compactified extra dimensions ($d$). The numerical integration was carried out using Python's function, {\tt scipy.integrate.solve\_ivp}, which employs the RK45 explicit Runge\,--\,Kutta method of order 5(4).
\begin{figure}[!ht]
     \centering     \includegraphics[scale=0.55]{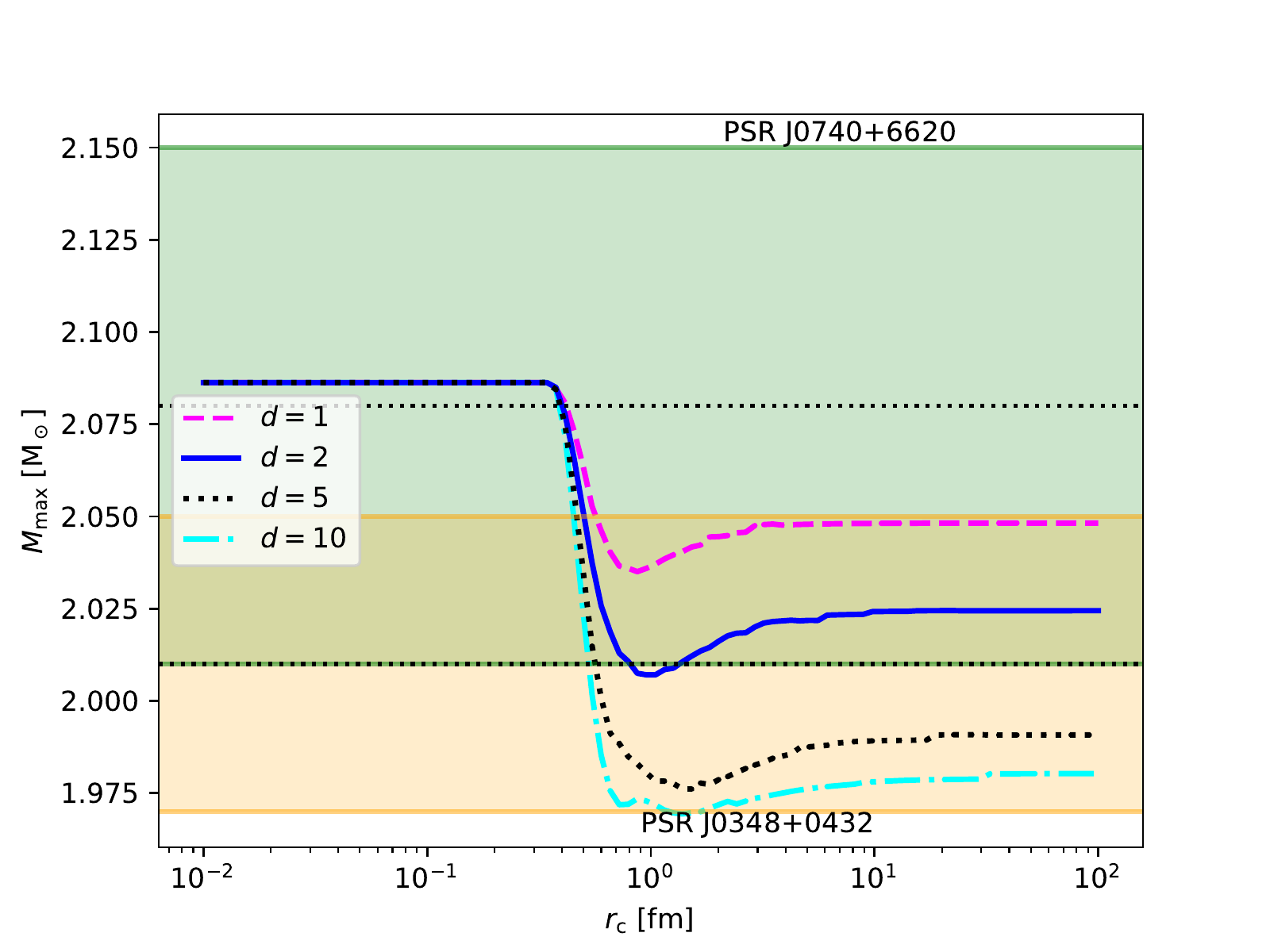}
     \caption{The maximal mass of compact stars plotted as a function of $r_{\mathrm{c}}$ calculated for equations of states with different parameter sets. Colors and line styles represent EoSs with different number of compact extra dimensions. The maximal excitation number was fixed to $N_\mathrm{exc}=1$, while the repulsive interaction strength to $\xi = 1000$~MeV$\cdot$fm\textsuperscript{3}. Colored bands correspond to uncertainties of pulsar mass measurements of J0740+6620~\cite{Fonseca_2021} and J0348+0432~\cite{Antoniadis:2013pzd}, which are likely to exhibit close to maximal mass. Horizontal dotted lines indicate the measurements' most probable values.}
     \label{fig:flock}
\end{figure}

We present our results in Fig.~\ref{fig:flock}, where the maximal mass of compact stars is plotted as a function of the size of the extra dimensions, $r_{\mathrm{c}}$.  Curves with different colors and line styles correspond to different numbers of extra dimensions, $d=1,$ 2, 5 and 10. The maximal excitation number was set to $N_\mathrm{exc}=1$, while the interaction strength to $\xi = 1000$~MeV$\cdot$fm\textsuperscript{3}, which is in the same order of magnitude as nuclear model predictions~\cite{Horvath_2024,Zimanyi:1987bt}. With this choice, we were able to reproduce measurement data corresponding to neutron stars likely to exhibit close to maximal mass, pulsars J0740+6620 and J0348+0432. Their observed masses~\cite{Fonseca_2021,Antoniadis:2013pzd} are shown in horizontal black dotted lines with uncertainties indicated by colored bands. One can see that extra dimensions in compact stars can be tested only above size  $r_\mathrm{c}\approx0.2$~fm, since below the energy needed to produce excitions in a significant amount is not reached. With one allowed excitation in each dimension, the maximal mass is reduced by approximately 2-6\%, which is similar in magnitude to the effects of the excitation number variation.

\section{Discussion and Conclusions}
\label{sec:sum}

Compact stars within a Kaluza\,--\,Klein-like model were considered, where spacetime is extended by $d$ compactified extra spatial dimensions at each spatial point in the usual 1+3 dimensions. Known, standard model particles are allowed to propagate in these extra directions, thus gaining quantized momenta, that form part of an effective mass that we get when performing reduction to the 1+3 dimensional theory. In this work, we showed how a simple equation of state can be formulated for this type of multi-dimensional matter with a repulsive linear potential. We calculated its properties, and how it can be used in the Tolman\,--\,Oppenheimer\,--\,Volkoff equation -- which takes its usual form -- to model compact stars~\cite{Barnafoldi:2015wca,Karsai:2016wfx,Horvath_2024}.

We found that the number of extra compactified dimensions plays a very similar role as the excitation number: the strength of the effect was found to be at the level of a few percent in solar mass units for both cases. However, the value of $N_\mathrm{exc}$ and $d$ are not directly comparable due to their interplay, and the distinct influence they have on the number of available states, indeed, multiplicities. In a general case, following from the thermodynamic potential, Eqs.~\eqref{eq:termpot}, increasing $d$ has a more prominent effect, however, it is also more constrained from a physical point of view, since most beyond standard model theories do not expect the existence of more than six extra dimensions. Thus, the effect of increasing $N_\mathrm{exc}$ or $d$ can be considered quantitatively the same.

\section*{Acknowledgments}

Authors gratefully acknowledges the Hungarian National Research, Development and Innovation Office (NKFIH) under Contracts No. OTKA K135515 and K147131, No. NKFIH NEMZ\_KI-2022-00031, 2024-1.2.5-TÉT-2024-00022 and Wigner Scientific Computing Laboratory (WSCLAB, the former Wigner GPU Laboratory). Author A.H. is supported by NKFIH through the DKÖP program of the Doctoral School of Physics of Eötvös Loránd University and HUN-REN's Mobility fellowship with indentifiers KMP-2023/101 and KMP-2024/31. E.F-D. also received funding from the NKFIH excellence grant TKP2021-NKTA-64.

\bibliographystyle{ws-ijmpa}

\bibliography{multidimkk}

\end{document}